\documentclass[journal]{IEEEtran}

\usepackage{hyperref}

\ifCLASSINFOpdf
   \usepackage{graphicx}
\else
   \usepackage[dvips]{graphicx}
\fi

\usepackage{aas_macros}

\begin{document}
%
\title{Multidimensional simulations of magnetic field amplification and electron acceleration to near-energy equipartition with ions by a mildly relativistic quasi-parallel plasma collision.}


\author{Gareth~C.~Murphy,
        Mark~E.~Dieckmann,        
        and~Luke~{O'C.}~Drury
\thanks{Manuscript received December 1, 2009; revised July 12, 2010; accepted
July 26, 2010. This work was supported in part by the Science Foundation
Ireland under Grant 08/RFP/PHY1694 and in part by the Swedish
Vetenskapsr\aa det.}
\thanks{G. C. Murphy and L. {O'C}. Drury are with the Dublin Institute for Advanced
Studies, Dublin 2, Ireland (e-mail: gmurphy@cp.dias.ie; ld@cp.dias.ie).}
\thanks{M. E. Dieckmann is with the Department of Science and Technology,
Linkšping University, 60174 Norrk\"ping, Sweden (e-mail: Mark.E.
Dieckmann@itn.liu.se).}
\thanks{Color versions of one or more of the figures in this paper are available online
at http://ieeexplore.ieee.org.}
\thanks{Digital Object Identifier 10.1109/TPS.2010.2063442}
}


%


\maketitle

\begin{abstract}
The energetic electromagnetic eruptions observed during the prompt phase 
of gamma-ray bursts are attributed to synchrotron emissions. The internal 
shocks moving through the ultrarelativistic jet, which is ejected by an 
imploding supermassive star, are the likely source of this radiation.
Synchrotron emissions at the observed strength require the simultaneous 
presence of powerful magnetic fields and highly relativistic electrons. 
We explore with one and three-dimensional relativistic particle-in-cell 
simulations the transition layer of a shock, that evolves out of the 
collision of two plasma clouds at a speed 0.9c and in the presence of a 
quasi-parallel magnetic field. The cloud densities vary by a factor of 10. 
The number densities of ions and electrons in each cloud, which have the 
mass ratio 250, are equal. The peak Lorentz factor of the electrons is 
determined in the 1D simulation, as well as the orientation and the strength 
of the magnetic field at the boundary of the two colliding clouds. The 
relativistic masses of the electrons and ions close to the shock transition 
layer are comparable as in previous work. The 3D simulation shows  
rapid and strong plasma filamentation behind the transient precursor. The magnetic field component orthogonal to the initial field direction 
is amplified in both simulations to values that exceed those expected from 
the shock compression by over an order of magnitude. The forming shock is 
quasi-perpendicular due to this amplification. The simultaneous presence of 
highly relativistic electrons and strong magnetic fields will give rise to 
significant synchrotron emissions.
\end{abstract}



%
\IEEEpeerreviewmaketitle


Mildly relativistic plasma collisions are commonly found in astrophysical 
scenarios, in particularly strong supernovae \cite{Koyama:1995ys,Kulkarni:1998lh,Enomoto:2002rt,Bamba:2003vn}, in 
gamma ray bursts (GRBs) \cite{Piran:1999jt,Medvedev:1999sf,Lyutikov:2003fk,Fox:2006vs} and 

in the jets 
of microquasars \cite{Fender:1999fk,Kaiser:2000fj,Jamil:2010lr} and active galactic nuclei \cite{Rees:1978zr} \cite{Spada:2001mz} \cite{Sahayanathan:2005uq}.

The prompt emissions of the
ultrarelativistic GRBs are probably the most spectacular eruptions of
electromagnetic radiation in the universe, as they are observed across 
cosmological distances. Since the first observations of GRBs thousands 
have been detected. They all share a common signature of energetic 
radiation attributed to highly relativistic electrons and strong magnetic 
fields. 

The details of the underlying physical mechanism which causes the prompt emissions is still unknown. 
In particular, the method by which magnetic field is generated and sustained, and the means by which electrons are accelerated to ultrarelativistic speeds in sufficient numbers to be injected into the Fermi mechanism is still under debate.
It is therefore interesting and relevant to use numerical simulations to probe the behaviour of magnetised shocks to see if a robust mechanism to amplify the field self-consistently may be found.

While the first-order Fermi 
acceleration mechanism is certainly capable of accelerating particles 
\cite{Drury:1983sf} to ultrarelativistic energies, it requires a seed 
population of non-thermal particles before it can viably accelerate them
to high energies. This mechanism may also not be able to provide rapidly
enough the huge quantities of highly relativistic electrons, which drive
the prompt emissions of gamma ray bursts. Plasma instabilities driven
by the relativistic flow may be a more likely candidate mechanism.

\subsection{Plasma Instabilities}

Many instabilities, such as the two-stream, filamentation or Weibel-type 
and oblique instabilities are expected to operate in the relativistic 
plasma, and it is not a straightforward task to distinguish the dominant 
mode \cite{Bret:2009ec},\cite{Lemoine:2009kg}. The Weibel-type filamentation 
instability is expected to generate or amplify magnetic fields. Since 
the pioneering studies of the electromagnetic Weibel instability 
\cite{Weibel:1959it,Morse:1971} and of the beam-Weibel or filamentation 
instability \cite{Davidson:1972,Lee:1973,Molvig:1975} much work has been 
done, both analytical \cite{Sagdeev:1979ph,Bret:2004dn,Bret:2005hl,Lazar:2009qq} 
and numerical using kinetic particle-in-cell simulations 
\cite{Honda:2000,Silva:2002qc,Stockem:2008hl} and Vlasov simulations 
\cite{Califano:2002ci}.

Reconnection in pair plasmas also excites the Weibel instability
\cite{Zenitani:2008qy,Swisdak:2008fj}.

Three-dimensional simulations have the best chance to approximate the correct 
physics, however they are extremely challenging even for contemporary state of 
the art supercomputers. It is thus customary to reduce the ion to electron mass 
ratio, to consider only leptonic flows or to extract aspects of the physically 
correct three-dimensional plasma behaviour and to model them in 1D or 2D grids 
at the appropriate spatio-temporal resolution. Certain instabilities can be 
examined in such reduced geometries like the electrostatic two-stream instability 
in one dimension and the filamentation instability in two dimensions.

\subsection{Previous related work on plasma collisions and shock formation}
Early work was done with counterstreaming and interpenetrating electron-positron 
plasmas, finding filamentation growing\cite{Silva:2002qc}\cite{Silva:2003rw}. 
Previous simulation studies of relativistic shocks in more than one spatial 
dimensions have focussed on mildly to highly relativistic collision speeds and
on symmetric clouds of positron-electron pair plasmas\cite{Kazimura:1998}
\cite{Spitkovsky:2005hi}\cite{Sironi:2009kt}\cite{Nishikawa:2009}. These simulations 
are fast, because they only involve one spatiotemporal scale, namely the leptonic 
one. However these scenarios necessarily exclude the wealth of wave modes and nonlinear
processes which exist in ion-electron collisions, due to the mass asymmetry and the 
massive kinetic energy of the ions. Relevant PIC simulations are the 3D simulation 
with a low mass ratio discussed in Ref. \cite{Frederiksen:2004rm} and the 2D simulation 
with a high mass ratio performed in Ref. \cite{Spitkovsky:2008Ion}. Magnetic field 
effects on the collision of two plasma clouds have been investigated too 
\cite{Hededal:2005zl}.

In this paper we consider the simultaneous effects of a high plasma temperature, a
quasi-parallel magnetic field and an initial density asymmetry on the plasma collision. 
A density asymmetry has been introduced before \cite{Frederiksen:2004rm}, but here we 
increase the density ratio from 3 to 10. This high ratio implies, that the initial 
spectrum of unstable waves should shift from the electromagnetic filamentation 
instability to the partially electrostatic oblique mode instability \cite{BreGrem:2007}. 
This shift is further emphasized by the high temperature and the guiding magnetic field, 
which both reduce the growth rate of the filamentation instability compared to those of 
the other waves. Section 1 provides a more detailed overview over the simulation code
and the initial conditions. One-dimensional PIC simulations with a similar setup have 
been considered first by Ref. \cite{Bessho:1999sf}. They have been extended to two 
dimensions, albeit over a limited spatio-temporal range, by Ref. \cite{Dieckmann:2008dp}. 
The 1D PIC simulation data we present here in Section 2 employs a setup that is 
practically identical to that in Ref. \cite{Dieckmann:2008dp} apart from the different ion 
to electron mass ratio and code. The simulation results are essentially the same, which 
is clear evidence for the actual physical equipartition of the electron and ion energy 
in the resulting shock. Obtaining this result only for one mass ratio may be a 
coincidence. The 3D numerical simulation also presented in Section 2 extends the 2D 
simulation in Ref. \cite{Dieckmann:2008dp}, confirming that the front of the dense cloud 
is practically planar. The 1D PIC simulation is thus at least initially a good 
approximation for the plasma dynamics close to the front of the expanding dense cloud. 
However, the plasma density distribution clearly evidences a filamentation behind this 
front, which is not revealed to this extent by the electromagnetic fields in Ref.
\cite{Dieckmann:2008dp}.

\section{Initial conditions and numerical Method}

Two plasma clouds collide in the simulation box. The species 1 and 2 are the electrons 
and ions of the dense cloud, while the electrons and ions of the tenuous cloud are the 
species 3 and 4. The species number gives the subscript $j$ of the density $n_j$. Each 
cloud has electrons and ions with the mass ratio $R=m_i / m_e = 250$, with the same 
density ($n_1 = n_2$ and $n_3=n_4$), temperature and mean speed. Each cloud is thus 
initially charge- and current neutral. The mean velocity vectors of both clouds point 
along opposite $x$-directions and they have the same modulus $v_b$. Each cloud occupies 
initially one half of the simulation box and their contact boundary is located at $x=0$ 
at the time $t=0$. Both clouds are uniform along the $y,z$ directions. No particles 
are introduced at the boundaries for $t>0$. The clouds thus detach instantly from the 
walls and we can use periodic boundary conditions in all directions. A guiding magnetic 
field $\mathbf{B}_0$ gives an electron gyrofrequency $\omega_c = e|\mathbf{B}_0| / m_e$ that 
equals the electron plasma frequency $\omega_{p1} = {(e^2 n_1 / m_e \epsilon_0)}^{1/2}$ of 
the dense cloud. The $\mathbf{B}_0$ is quasi-parallel to the flow (x) direction and at 
$t=0$ the $B_{0,x} \gg B_{0,z}$ and $B_{0,y}=0$. The convection electric field $E_c = 
|v_b B_{0,z}|$ changes its sign across the collision boundary. The collision speed $2v_b / 
(1+v_b^2) = 0.9c$ and the electron thermal velocity $v_t = {(T_e / m_e)}^{1/2}$ for $T_e =$ 
131 keV is $\approx$ c/2. We express space and time in the relevant units of the dense ions 
(species 2), which are the ion skin depth $\lambda_2 = c / \omega_{p2}$ with 
$\omega^2_{p1}/\omega^2_{p2}=R$ and the inverse ion plasma frequency $\omega_{p2}^{-1}$.

We use the particle-in-cell method described in detail in Ref. \cite{Dawson:1983dz}.
The normalised equations are
\begin{equation}
\nabla  \times \mathbf{E} =  -\frac{\partial \mathbf{B}} {\partial t},
\end{equation}
\begin{equation}
\nabla  \times \mathbf{B} =  \frac{ \partial \mathbf{E}} {\partial t} + \mathbf{J},
\end{equation}
\begin{equation}
\nabla  \cdot \mathbf{B} =  0,
\end{equation}
\begin{equation}
\nabla  \cdot \mathbf{E} =  \rho,
\end{equation}
\begin{equation}
\frac{ d \mathbf{p}_j}{d t} = q_i \left( \mathbf{E} + \mathbf{v}_j \times \mathbf{B}
\right),
\end{equation}
\begin{equation}
\mathbf{p}_j=m_k \Gamma_j \mathbf{v}_j,
\end{equation}
\begin{equation}
\frac{ d \mathbf{x}_j}{d t} =\mathbf{v}_j,
\end{equation}

The particle $j$ of species $k$ has the mass $m_k$.
The quantities in SI units (subscript $p$) can be obtained by the substitutions 
$\mathbf{E}_p = \omega_{p2} c m_i \mathbf{E} / e$, $\mathbf{B}_p = \omega_{p2} m_i 
\mathbf{B} / e$, $\rho_p = e n_2 \rho$, $\mathbf{J}_p = e c n_2 \mathbf{J}$, 
$\mathbf{x}_p = \lambda_2 \mathbf{x}$ and $t_p = t / \omega_{p2}$.

The simulation resolution is as follows. For the 1D simulation we use 18,000 cells 
along $x$ and 250 particles per cell. For the 3D simulation we can only simulate a 
small fraction of the 1D domain with regard to $x$, but we get instead a view of 
the early 3D filament formation. We use 20 particles per cell and a 3D box composed 
of 1500x100x100 cells, which spans a total of $15\lambda_2 \times 1.25\lambda_2 \times 
1.25\lambda_2$ and a long direction aligned with $x$. 
The timesteps for the 1D and 3d simulations are both $8.6e2 \omega^{-1}_{p1}$
The physical and simulation 
parameters are summarized in Tables \ref{t1} and \ref{t2}.

\begin{table}[!t] 
\renewcommand{\arraystretch}{1.3} 
\caption{Physical Parameters} 
\label{t1} 
\centering 
\begin{tabular}{c||c} 
\hline 
\bfseries Parameter & \bfseries Value\\ 
\hline\hline 
Thermal Velocity & $0.52c$\\ 
Collision Speed & $0.9c$\\ 
Beam Speed & $0.63c$ \\
Temperature & 131 keV \\
Mass ratio & 250 \\
Field Angle &  0.1 radians\\
Density Ratio & 10 \\
\hline 
\end{tabular} 
\end{table} 

\begin{table}[!t] 
\renewcommand{\arraystretch}{1.3} 
\caption{Numerical Parameters} 
\label{t2} 
\centering 
\begin{tabular}{c||lc} 
\hline 
\bfseries Parameter & \bfseries 1D & \bfseries 3D \\ 
\hline\hline 
No. particles/cell & 250 & 20  \\ 
No. Cells& 18000 & $1500x100x100$ \\ 
Timestep &  8.6e-2 $\omega^{-1}_{p1}$ & 8.6e-2 $\omega^{-1}_{p1}$ \\
\hline 
\end{tabular} 
\end{table} 

\section{Results}

\subsection{One-dimensional simulation}

\begin{figure*}[t]
\begin{center}
   \begin{minipage}[t]{.48\linewidth}
\includegraphics[width=.99\linewidth]{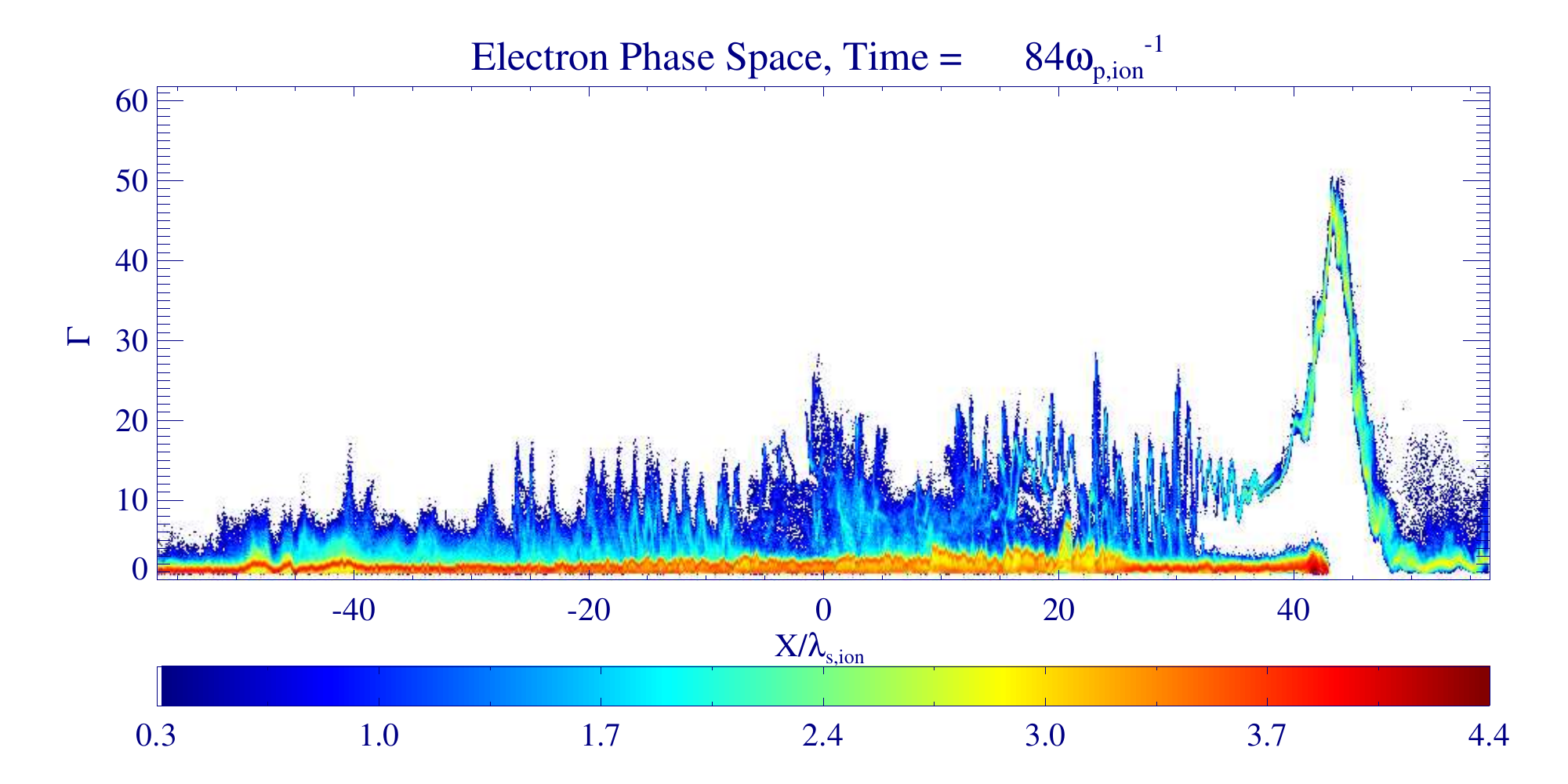}
   \end{minipage} \hfill
   \begin{minipage}[t]{.48\linewidth}
\includegraphics[width=.99\linewidth]{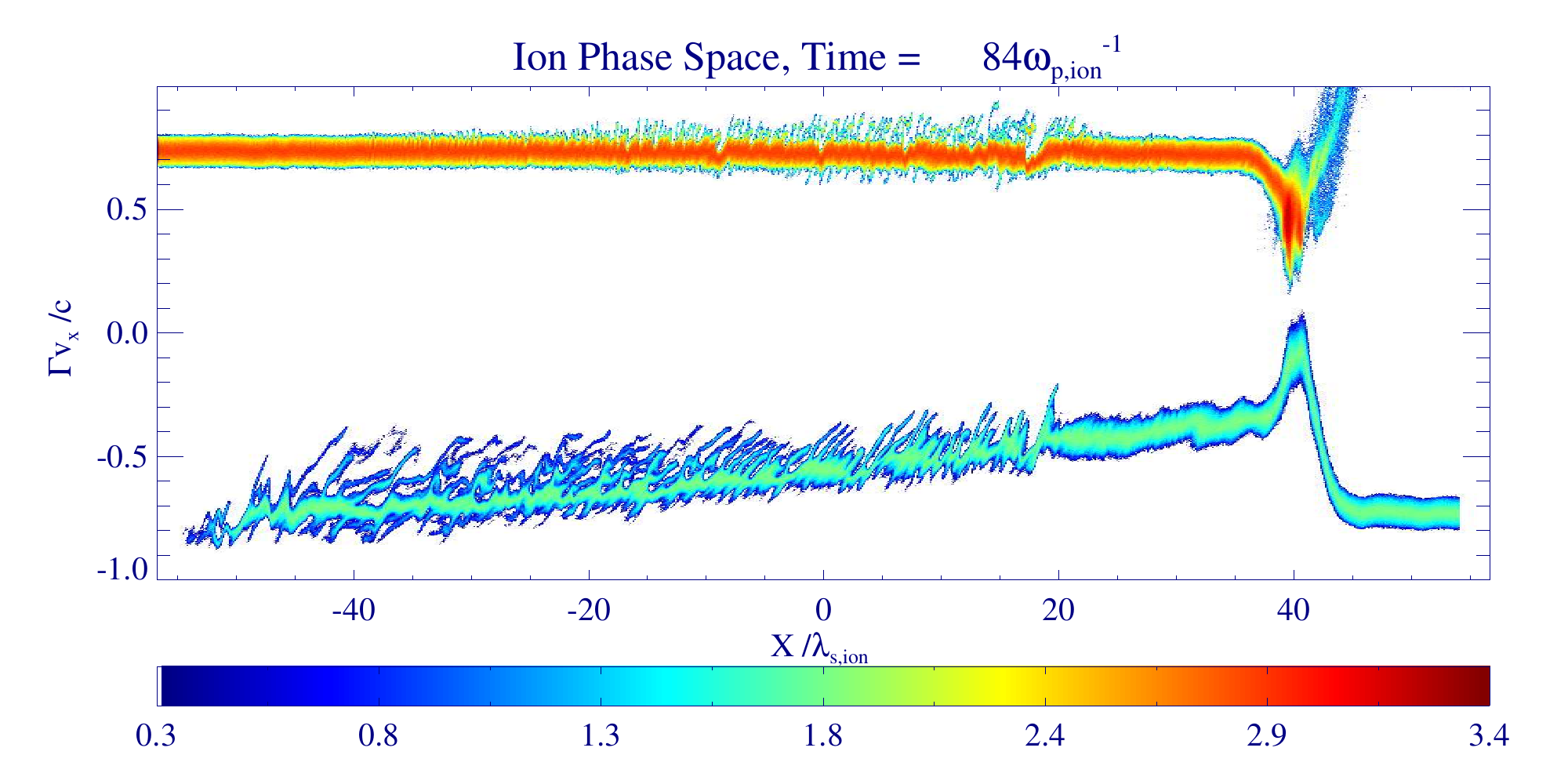}
   \end{minipage} \hfill
\hfill
\caption{Left Panel: Electron phase space density $f_e(\Gamma,x)$. Right Panel: Ion phase space density $f_i (\Gamma v_x/c,x)$. The colour scale is 10-logarithmic and the density is expressed in units of a computational particle of the tenuous cloud.}
\label{f1}
\end{center}
\end{figure*}

\begin{figure}[pb]
\centerline{\includegraphics[width=.99\linewidth]{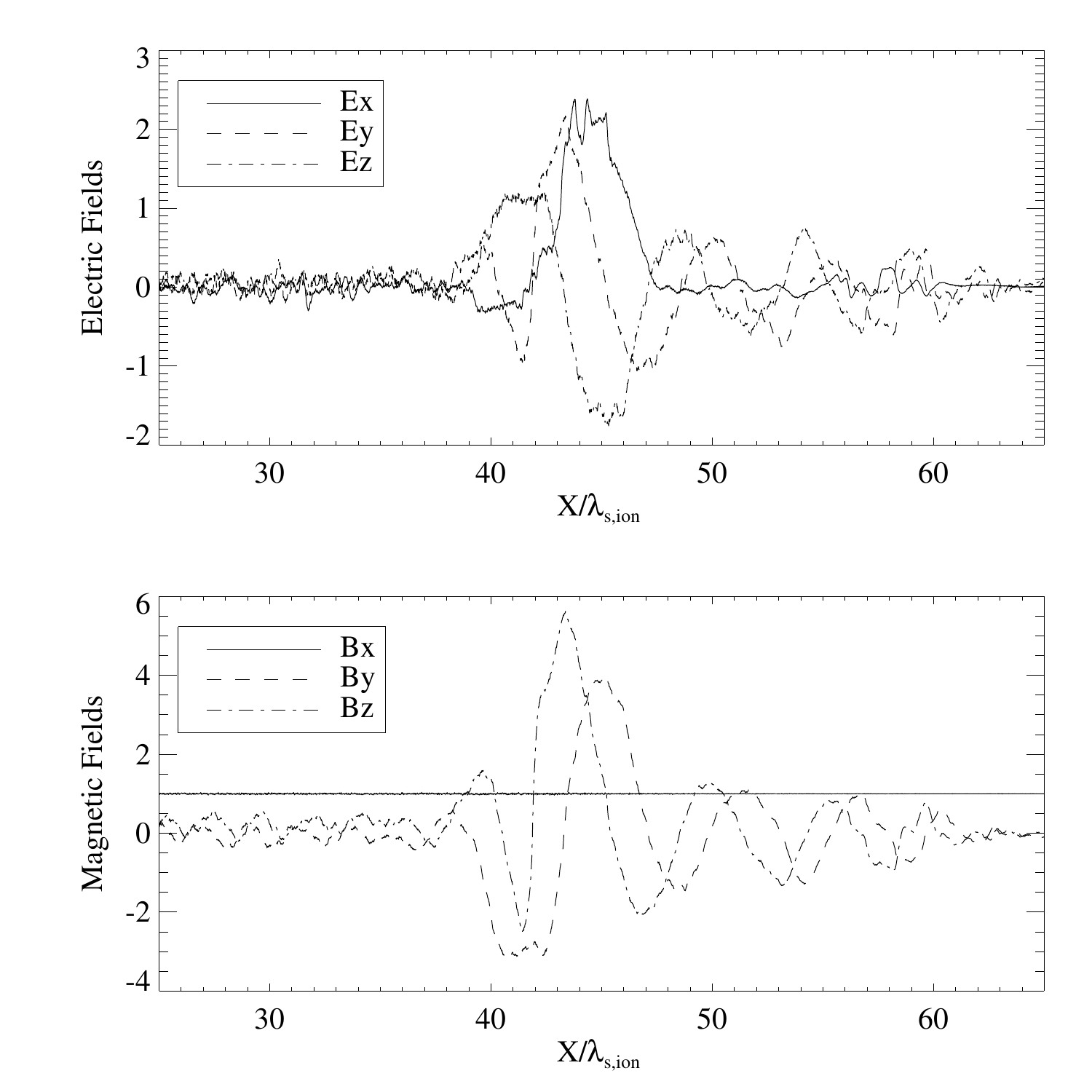}}
\vspace*{8pt}
\caption{The Energetic Electromagnetic Structure in 1D: Plots of electric and magnetic field strengths. The upper panel shows the components of $\mathbf{E}$ and the lower those of $\mathbf{B}$. }
\label{f1a}
\end{figure}

A circularly polarized electromagnetic wave grows practically instantly at the contact boundary
of both clouds and in both simulations. This collision boundary evolves due to the plasma
counterstream into a cloud overlap layer. The magnetic field is frozen-in into each cloud outside
the overlap layer. Inside this layer, the magnetic field continues to be practically at rest in
the reference frame of the dense plasma cloud. The tenuous cloud moves at a mildly relativistic 
speed relative to this reference frame and the particles are deflected by $\mathbf{B}_0$. The 
ions and electrons react differently to $\mathbf{B}_0$ and, due to its obliqueness, the particles 
are forced onto an orbit that involves all components of $\mathbf{p}$. This corkscrew orbit is 
discussed in more detail elsewhere \cite{Dieckmann:2008dp}. The resulting net current amplifies the 
magnetic field perturbation and a circularly polarized localized wave structure forms. 
  
Figure \ref{f1} displays the most relevant phase space distributions. These are the electron
distribution $f_e (\Gamma,x)$ and the ion distribution $f_i (x,p_x)$. The electron distribution
reveals the peak energies reached by the electrons, while the ion distribution can reveal the
formation of a shock. This is achieved, when the ion distributions of both clouds mix in this
phase space projection. The 1D simulation stops at the time $t=84$, just before the shock 
formation. The electron distribution demonstrates that the electrons of the tenuous cloud
(species 3) are forced onto a strong oscillation, which resembles a corkscrew orbit by the
rotation in the $p_y,p_z$-plane. They reach a peak $\Gamma \approx 50$ at $x\approx 43$. The 
electrons of the dense cloud (species 1), on the other hand, are not yet accelerated on this 
scale. The electrons of both clouds show a two-stream configuration in the interval $30<x<43$. 
The electrons would drive the oblique mode instability, which is here suppressed by the 1D geometry. 
Only the electrostatic two-stream instability can develop here and it results in the thermalization 
of the electrons of the tenuous cloud. This thermalization results in the hot electron population 
with $x<30$ and with $\Gamma \le 25$. The phase speed of the two-stream waves is close to the beam 
speed of the tenuous beam and only those interact resonantly with this wave. The electron phase 
space distribution of the dense cloud thus remains practically unchanged. The ion phase space 
distribution reveals that a shock is about to form at $x\approx 40$. The circularly polarized 
energetic electromagnetic structure (EES) is here strong enough to force the ions onto a corkscrew 
orbit. In particular the ions of the tenuous cloud (species 4) are perturbed within the cloud 
overlap layer. The beam striation in the $x,p_x$ plane implies, that the associated fields must 
be at least partially electrostatic. Electrostatic fields and such phase space striation are 
reminiscent of the Buneman instability, which would develop here between the ions of the tenuous 
cloud and the electrons of the dense one \cite{Buneman:1958}.

Figure \ref{f1a} displays the distributions of all components of $\mathbf{B}$ and 
$\mathbf{E}$ near the front of the dense cloud at the time $t=84$, which corresponds to 
the particle distribution in Fig. \ref{f1}. The upper panel shows that all components of 
$\mathbf{E}$ have reached a comparable strength. The electromagnetic $E_y$ and $E_z$ are 
phase-shifted by $90^\circ$. The magnetic $B_y,B_z$ components (lower panel) also show 
this phase shift. We find that $B_x = B_{x,0}$. The $B_x$ must remain constant in a 1D simulation 
by $\nabla \cdot \mathbf{B} = 0$ and a change of $B_x$ in time could only be driven by $\partial_y 
E_z - \partial_z E_y \neq 0$, which is not possible in 1D. The asymmetry in space of the EES 
relative to its amplitude maximum may reflect the varying plasma skin depth across the collision 
boundary. The skin depth is reduced by a factor $\sqrt{10}$ in the dense plasma. If the penetration 
depth of the EES is a few ion skin depths, then the wave envelope must decrease more quickly as 
it enters the dense plasma.

The electrostatic component, $E_x$ is strong in the interval where the electromagnetic fields ($E_y$ and $E_z$) are strong. 
The electromagnetic fields are tied to the circularly polarized wave structure, which provides 
the dissipation of the flow energy. The $E_x$ is such, that electrons are accelerated to the 
left within $42 < x < 47$. This electric field must develop between the ions and the electrons 
of the tenuous cloud, because the dense cloud in Fig. \ref{f1} has not propagated yet that far. 
The likely cause is the deflection of the electrons of species 3 by the EES. A mere rotation of 
the electron velocity vector away from the $x$ direction, would decrease their flow velocity 
along $x$. The ion speed remains practically unaffected and an electric charge builds up. The
 charge results in an electrostatic field, which drags the electrons with the ions. The 
electrons are accelerated at the expense of the ion kinetic energy. This explains the decrease 
of the flow speed modulus of the ions of the tenuous cloud and the electron acceleration in 
this interval. 

The EES at the front of the dense cloud has a wavevector parallel to $x$ and it can thus be 
resolved by the 1D simulation. The electromagnetic instabilities in the cloud overlap layer, 
e.g. the filamentation- and the oblique mode instability, cannot develop here, because their 
wavevectors are orthogonal or oblique to the beam velocity vector. The plasma dynamics in the 
cloud overlap layer can thus not be accurately resolved by the 1D simulation, which overemphasizes 
the importance of the electrostatic two-stream and Buneman instabilities with their flow-aligned
wavevectors. We turn to the 3D simulation, with which we can resolve the full wave spectrum
during the first few inverse ion plasma frequencies.

\subsection{Three-dimensional plasma filament structure}

\begin{figure}[pb]
\centerline{\includegraphics[width=.99\linewidth]{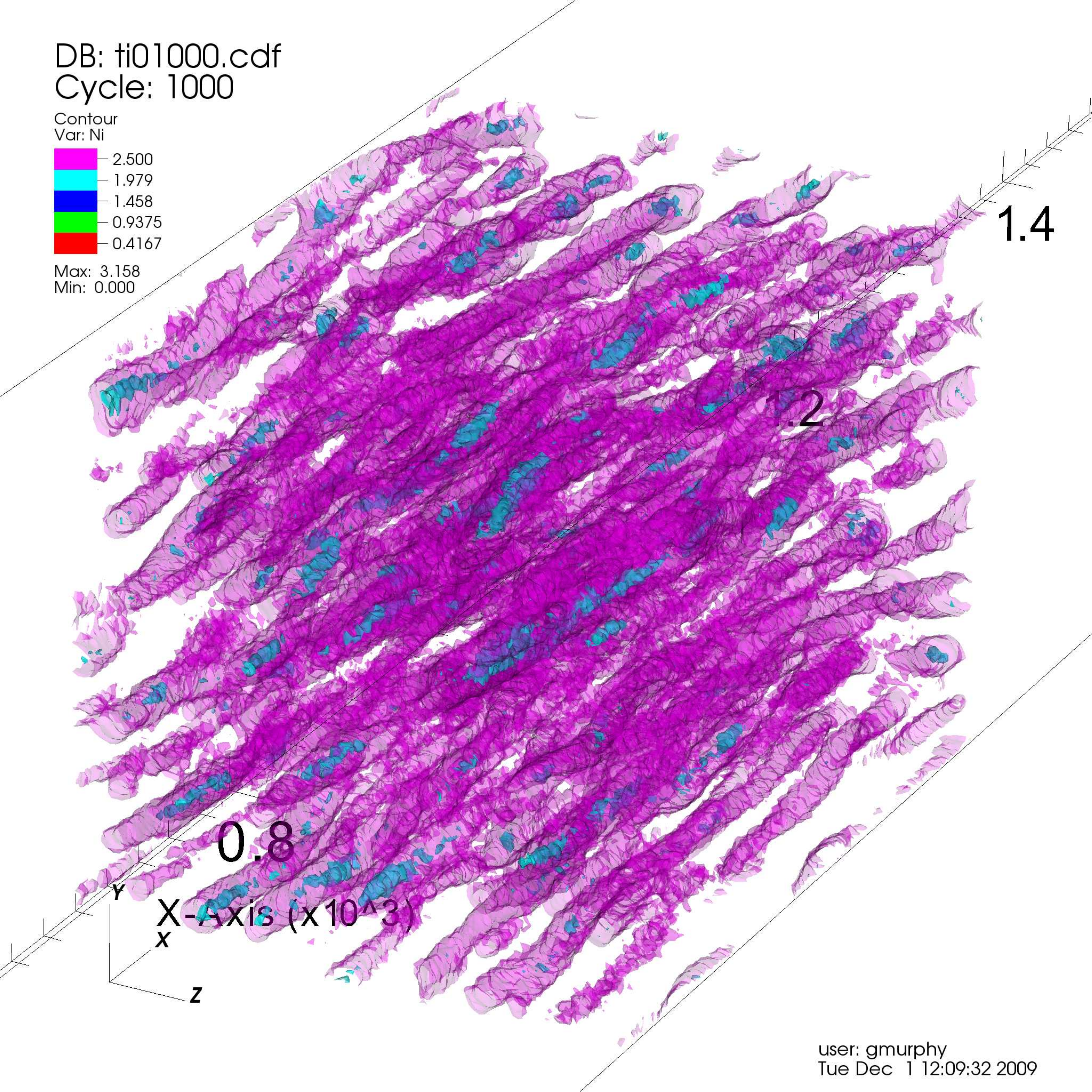}}
\vspace*{8pt}
\caption{Three-dimensional rendering of a simulation of plasma collision. Two isosurfaces of logarithm of ion density are shown at values of 2.5 and 1.9. The filaments are preferentially aligned along the x axis, gyrating in the y z plane.    }
\label{f2}
\end{figure}

\begin{figure*}[t]
\begin{center}
   \begin{minipage}[t]{.48\linewidth}
\includegraphics[width=.99\linewidth]{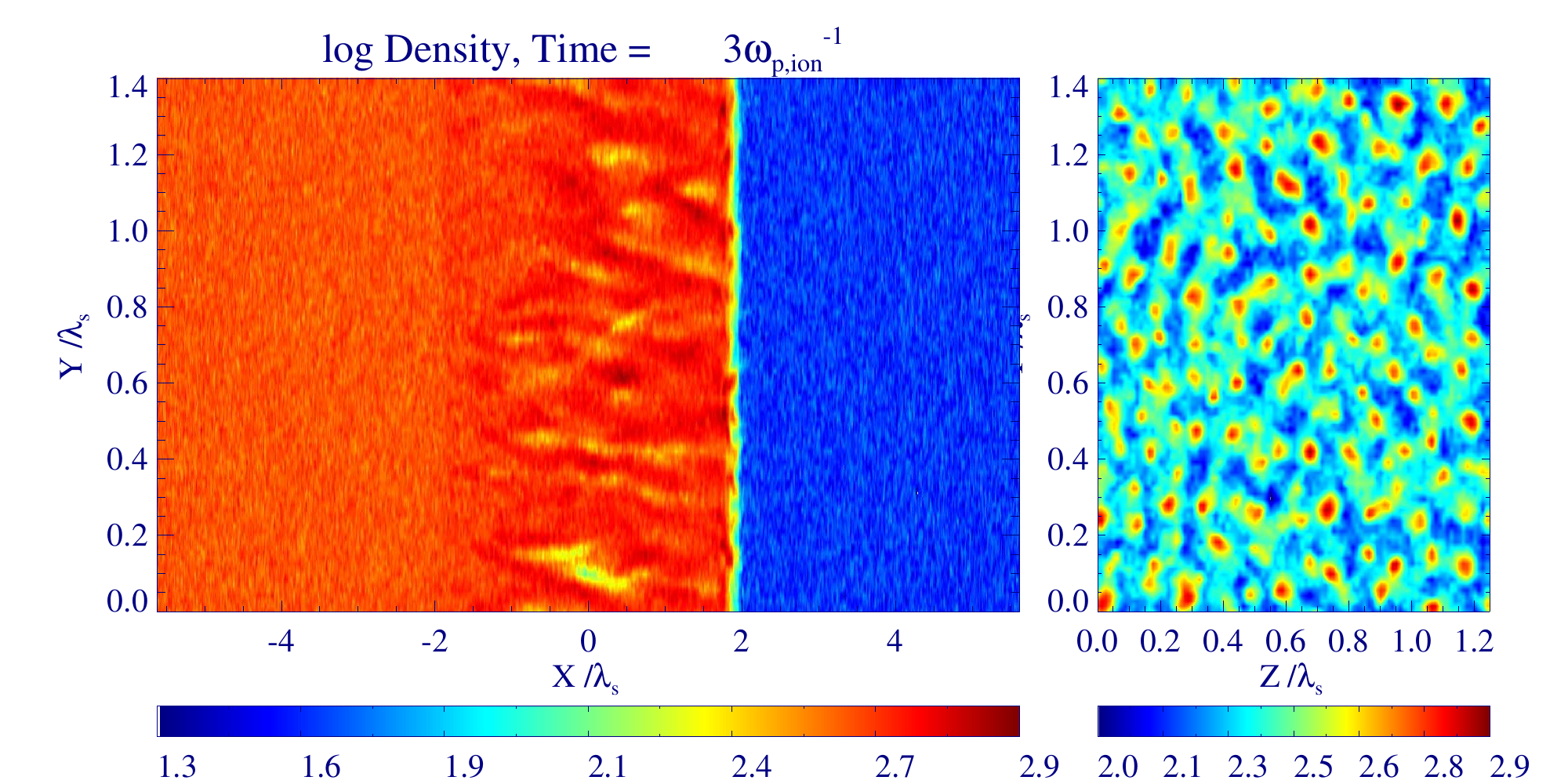}
   \end{minipage} \hfill
   \begin{minipage}[t]{.48\linewidth}
\includegraphics[width=.99\linewidth]{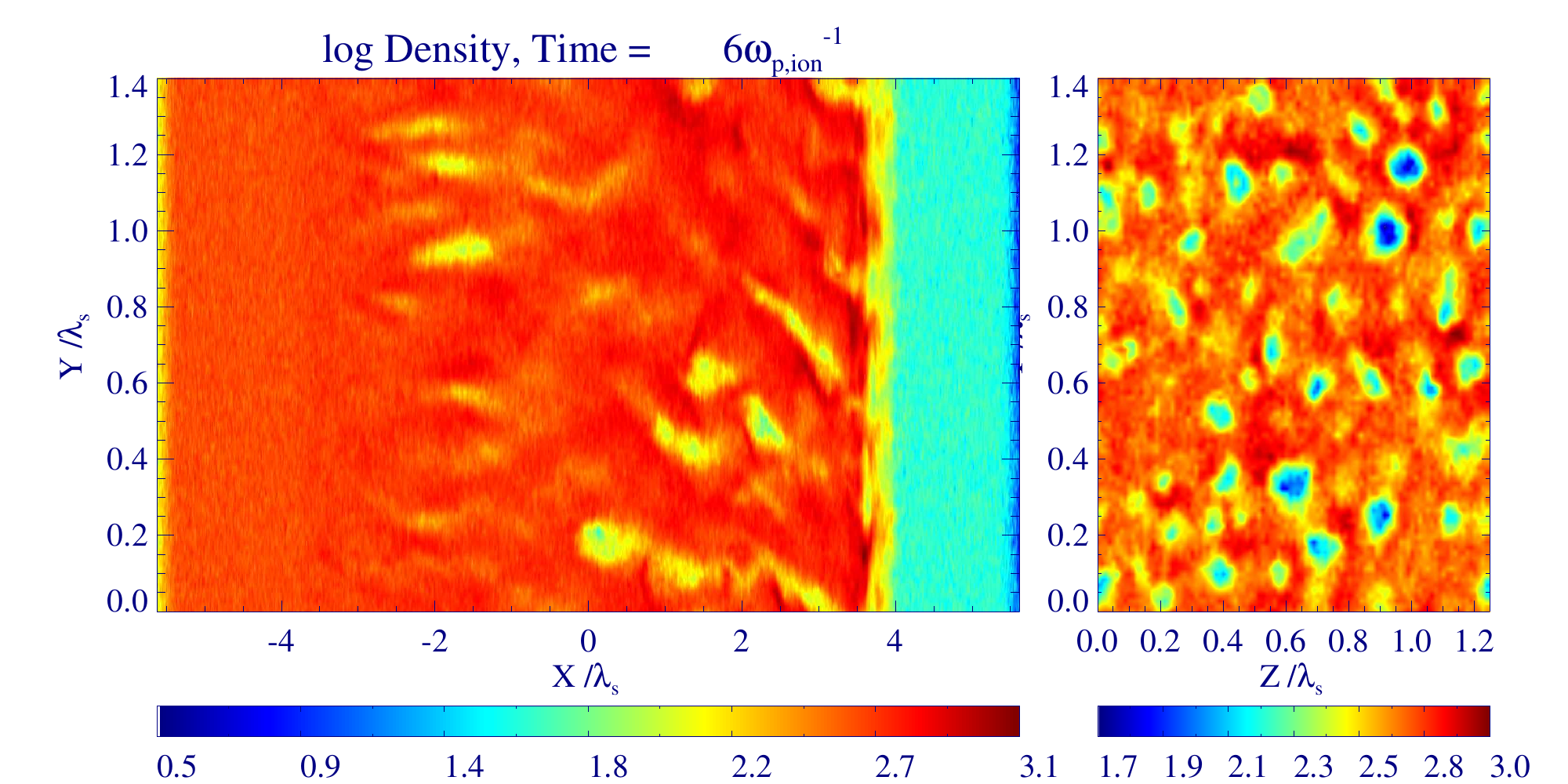}
   \end{minipage} \hfill
\hfill
\caption{Growth and merging of filaments in 3D: Ion density plots at time $t=3$ and $t=6$. 
Filaments, which are characterized through density modulations, have increased in size.}
\label{f3}
\end{center}
\end{figure*}

\begin{figure}[pb]
\centerline{\includegraphics[width=.99\linewidth]{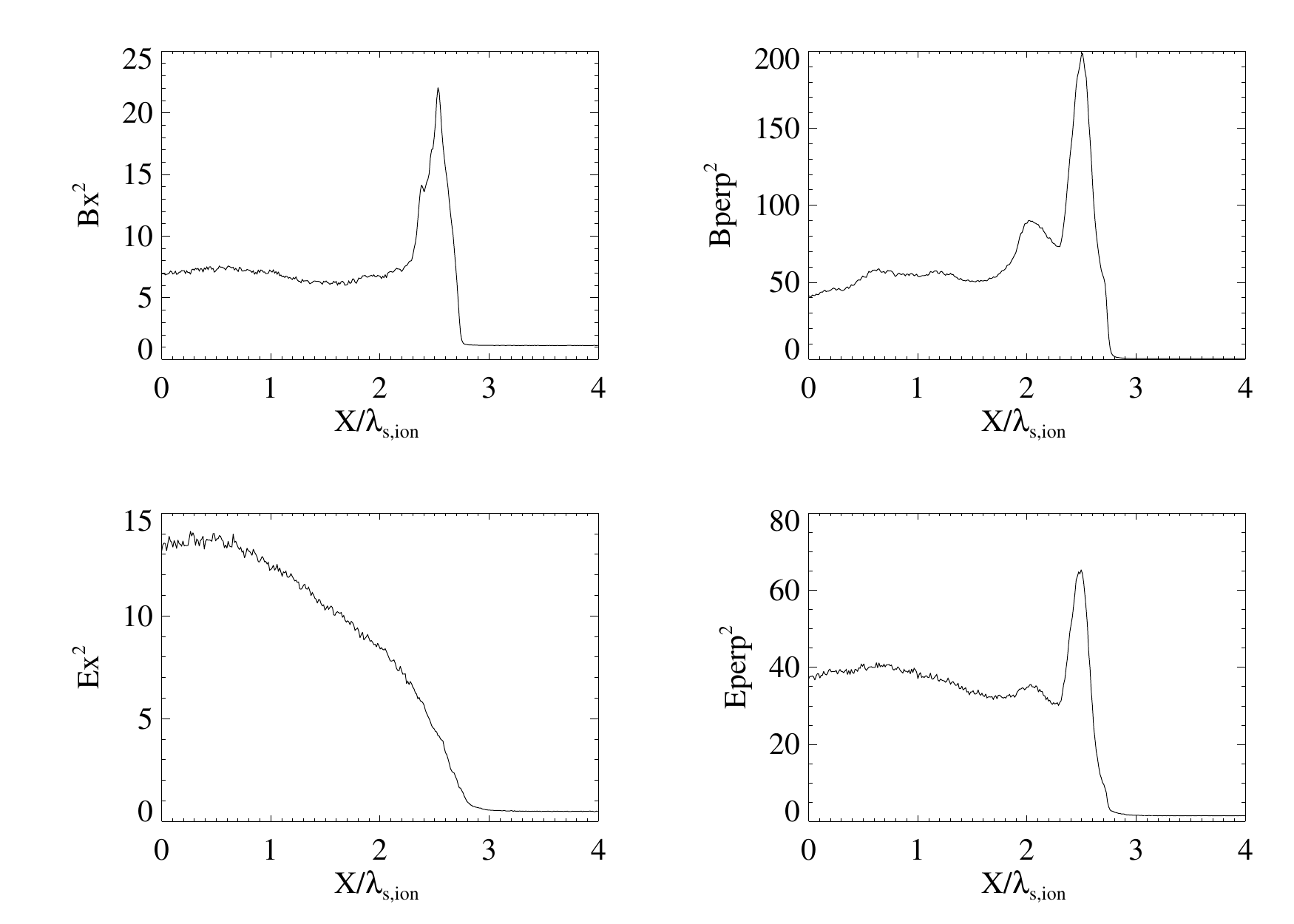}}
\vspace*{8pt}
\caption{Electromagnetic energy densities at $t=3$. The left panels show the energy densities
$B_x^2$ and $E_x^2$ along the initial flow velocity vector, while the right panels show
$B_{perp}^2 = B_y^2 + B_z^2$ and $E_{perp}^2 = E_y^2 + E_z^2$ orthogonal to the initial flow 
direction.}\label{f4}
\end{figure}

The 3D simulation is initialised in the same manner as the 1D simulation, except over a smaller 
domain along $x$, due to computational contraints. Consequently the simulation is run over the 
significantly shorter time interval $t=6$ than the 1D simulation with its $t=84$. This time is 
also somewhat shorter than the runtime of the 2D PIC simulation in Ref. \cite{Dieckmann:2008dp}, 
which ran for about 11 inverse ion plasma frequencies for the mass ratio 400. This previous work 
examined the electromagnetic wave structures in particular at the front of the dense cloud and 
the simulation's end, showing that they have remained planar. Here we do not look at the field 
topology but at the ion density distribution in the entire simulation box. 
  
Figure \ref{f2} shows a 3d rendering of the ion density in our numerical simulation. Clearly 
the counterstreaming ions in three dimensions quickly form filaments, which result in an 
amplification of the magnetic field orthogonal to the filament axis. These magnetic fields
grow until they saturate through magnetic trapping \cite{Davidson:1972}. The strength $B_0$ 
of the guiding magnetic field has been selected so as to suppress the filamentation 
instability and the oblique mode instability driven by the electrons \cite{Bret:2006rc} in 
a spatially uniform two-beam configuration. The magnetic field strength has apparently 
been insufficient to suppress the filamentation and mixed mode instabilities in the presence 
of ions and of a spatial nonuniformity. A further increase of the amplitude of $B_0$ would 
be necessary to achieve this \cite{Hededal:2005zl}, but such a field is probably unrealistic 
for most astrophysical plasmas. The filaments have a preferential alignment with the initial 
flow velocity vector. This is not surprising, because the filamentation and oblique mode 
instabilities yield the growth of a magnetic field through the redistribution of the 
microcurrents carried by the moving charged particles. These microcurrents $\propto q \mathbf{v}$ 
are obviously strongest along the mean flow (x) direction. 

This preferential alignment of the filaments with the x-direction allows us to obtain insight 
into the filament thickness with slices in the y,z plane. Their coherence length along the 
filament axis can be estimated from two-dimensional cross-sections in the x,y plane. In Figure 
\ref{f3} we plot these two dimensional cuts from the 3d simulation at the times $t=3$ and $t=6$. 
We clearly see the formation of filaments in the ion density distribution, which we identify 
through a density modulation. Their characteristic diameter is well below 0.1 ion inertial 
lengths at $t=3$ and is thus comparable to the electron skin depth of the dense cloud $\lambda_1 
= \lambda_2 / \sqrt{R}$. The filament diameter increases beyond 0.1 ion inertial lengths through 
the dynamical interaction and the mergers of filaments \cite{Bogdan:1985}. This small size 
suggests that until $t=6$ the main driver of the ion beam filamentation has been the electrons. 
The cut of the ion distribution in the y,z plane in the right panel of Fig. \ref{f3} shows some 
filaments with a significant density depletion and with a radial cross-section. Examples are the 
low-density structures at $z \approx 1.0$ and $0.9 < y < 1.2$. These are ion filaments immersed 
in an almost spatially uniform ion background. The ion density is reduced and the small spatial 
scale suggests that the ions merely follow the electron density, in order to maintain the
quasi-neutrality of the plasma. The electron density can be decreased through a magnetic expulsion 
of the electrons of one of the clouds. The coherence length in the x,y plane is about an ion skin 
depth, which is an order of magnitude larger than the filament diameter. Some of these filaments 
may thus be described well by their cross section in the y,z plane, which neglects changes along 
$x$. We furthermore find that the spatial separation of these radial filaments is larger than 
their diameter, which implies that these filaments may not interact magnetically over a limited 
time interval. An equilibrium might be possible during this time between the filament and the 
spatially uniform background plasma. This equilibrium may be similar to the Hammer-Rostoker 
equilibrium for a tenuous relativistic electron beam that crosses a dense electron background 
\cite{Hammer:1970}. The agreement is not exact. Here the ions react to the electric charge built 
up by the expulsion of the background electrons from the flux tube and the plasma carries an 
oblique magnetic field. Both aspects have not been considered by Ref. \cite{Hammer:1970}. 
 
Finally we notice from the right panel in Fig. \ref{f3} that the filamentation also involves
the front of the dense cloud at $x\approx 4$. This filamentation is fairly weak and its scale
is again a fraction of the ion skin depth. This implies that a 1D simulation can not reproduce
the exact physics at the cloud front but, at least until $t=6$, this approximation is fairly
accurate. It is interesting that a filamentation was not observed in the 2D simulation in
Ref. \cite{Dieckmann:2008dp} that covered a longer time interval. Several reasons are possible. 
Firstly, the higher ion mass in the 2D simulation may have delayed the ion filamentation in
the y,z plane with respect to the formation of the EES, which modulates the ions along $x$. Both 
are competing processes. Secondly, the field distribution was used in Ref. \cite{Dieckmann:2008dp} 
to determine the planarity of the cloud front, while the ion density is considered here. The
magnetic field amplification by the filamentation instability is much weaker than that by
the EES and it may not be visible. The ion density modulation along y,z can thus more accurately
reveal the filamentation than the electromagnetic fields. The electromagnetic fields due to the
EES reach a much higher energy density than those due to the filamentation instability also in
the 3D simulation, which we show now.

\subsection{The  Energetic Electromagnetic Structure (EES) in 3D}

The EES acts to accelerate charged particles, it can provide the energy dissipation that 
enforces the formation of a shock and it may provide a compact location for synchrotron 
radiation in the shock transition layer. The size scale and the energy density is similar 
to that detected in the Short Large Amplitude Magnetised Structures (SLAMS), which have 
been observed both in simulations and in those regions of the Earth's bow shock, where the 
ambient magnetic field is quasi-parallel to the shock normal \cite{Dieckmann:2008dp}\cite{Mann:1994}\cite{Behlke:2003}
\cite{Lembege:2004}\cite{Dieckmann:2010qf}. 

In Figure \ref{f4}, we see for the first time in 3d dimensions the EES. We make use of the 
quasi-planarity of the front of the dense cloud at $t=3$ in Fig. \ref{f3} and we integrate
the energy densities of the electromagnetic fields along y and z, which reduces the noise 
levels. The energy densities subdivide naturally into (1) $B_x^2$, which is unchanged in a 
1D simulation, into (2) $B_{perp}^2 = B_y^2+B_z^2$ and (3) $E_{perp}^2 = E_y^2 + E_z^2$, which 
would be purely electromagnetic in a 1D simulation and into (4) $E_x^2$, which would be 
electrostatic. This distinction is strictly possible only in 1D, because the derivatives 
along y and z vanish in the Maxwell's equations. We expect that initially and due to the 
planarity of the wave structures this holds also in 3D. All energy densities are normalized 
to the same value, which is the energy density of $\mathbf{B}_0$. 

Already at this early simulation time all energy densites surpass by far the initial magnetic
energy density as the front of the dense plasma cloud is crossed from the upstream into the
cloud overlap layer. The dominant component is $B_{perp}$, as in the 1D simulation. It peaks
at $x\approx 2.6$, which is ahead of the dense cloud. The peak of $E_{perp}$ coincides with
this location and both thus belong to the same wave structure. Note that, even if the EES 
were purely magnetic in its rest frame, the convective electric field would be significant
due to its rapid propagation. The three-dimensional simulation facilitates the growth of
$B_x^2$ and its peak energy density, which is an order of magnitude less than that of 
$B_{perp}^2$, is reached at the same position $x=2.6$ as that of $B_{perp}^2$. The $E_x^2$
in the 3D simulation shows like that of $B_x^2$ a distribution, which differs strongly from 
its equivalent in the 1D simulation. It grows steadily as we go from $x=3$ to $x=0.5$. The
electric field in the 1D simulation peaked where the EES was strongest. We interpret the
changed distribution $E_x^2$ in the following way. Let us assume, that the $E_x$ is driven 
in 1D entirely by the different deflection of the electrons and the ions of the tenuous cloud. 
The quasi-neutrality of the plasma of the tenuous cloud can then only be maintained by a 
flow along x, while a 3D simulation enables also a flow of charges along y and z. This
will, in turn, alter the electric field distribution. The apparent consequence is that the
$E_x^2$ disappears within the EES. We may attribute instead the increase of $E_x^2$ in the
3D simulation to the particle heating. The energy density of the electrostatic fluctuations 
is proportional to the plasma temperature, if the plasma is in an equilibrium (See 
\cite{Dieckmann:2004} and references therein). Evidence for this connection is that $E_{perp}^2 
\approx 2E_x^2$ and the energy density per degree of freedom is thus similar. Since the 
electrostatic energy density in a thermal equilibrium is also proportional to the particle
mass, it is exaggerated by the low statistical plasma representation by a PIC code. The
energy density of the EES is, on the other hand, independent of the mass of a computational
particle and determined by the kinetic energy density of the inflowing plasma. We may expect 
that in reality the electromagnetic energy would far exceed the electrostatic one.

\section{Conclusions}
In this paper we presents the results of long term 1D and short-term 3D simulations of plasma collisions.
We found that in 1D a strong amplification of the field is found in the foreshock, in excess of that expected from compression alone. 
Strong circularly polarised electromagnetic waves are generated at the collision boundary.
We find that electron are accelerated to ultrarelativistic speeds and that the presence of the EES is confirmed in 3D
The quasi-parallel magnetic field fails to suppress the filamentation in three dimensions, which was not detected in 2D fields in previous work, probably obscured in the data due to the large contrast provided by the EES. Ion densities are found to be a much better indicator for current filaments.
The three dimensional structure of the EES differs from its one-dimensional model, in particular the parallel electric field is diminished at the collision boundary, and is only excited by electron heating.
Filaments are produced with a characteristic width of approximately 0.1 ion inertial lengths and a larger separation between filaments, indicating a possibility for a magnetised Hammer-Rostoker equilibrium. 
The presence and size of the EES could be critical for the acceleration of matter to high energies in plasma shocks.
The simulations show the relevance of three dimensional simulations, even over small physical scales. 

 The EES is relevant to the problem of the large fields and electron acclerations needed to explain the observations of spectra of Gamma Ray Bursts.
The study shows that the EES is stable in 3d and that the magnetic field is amplified in 3D.
Comparing other magnetic field amplification mechanisms such as Bell's instability 
\cite{Bell:2004eu}, this mechanism does not require addition of external current, but produces electron accleration to near equipartition speeds and a strongly amplified field self-consistently.
Shock compression can also amplify the field but only up to a factor $f$, $4<f<7$ depending on the value of the ratio of specific heats.
In this respect it is an important result.	

For future work, we need significantly longer timescales to confirm long-term stability of Energetic Electromagnetic Structure.
Two-dimensional simulations are already under way with the goal of estimating the long term evolution of the system at high resolution.

\section*{Acknowledgment}

The project is supported by Science Foundation Ireland grant number 08/RFP/PHY1694 and by the Swedish Vetenskapsr\aa det. The authors wish to acknowledge the SFI/HEA Irish Centre for High-End Computing (ICHEC) for the provision of computational facilities and support. The Plasma Simulation Code (PSC) was developed by Hartmut Ruhl at the University Ruhr, Bochum.



%
\bibliographystyle{IEEEtran}
\bibliography{IEEEabrv,MyBibDeskLibrary}

\begin{thebibliography}{10}
\providecommand{\url}[1]{#1}
\csname url@samestyle\endcsname
\providecommand{\newblock}{\relax}
\providecommand{\bibinfo}[2]{#2}
\providecommand{\BIBentrySTDinterwordspacing}{\spaceskip=0pt\relax}
\providecommand{\BIBentryALTinterwordstretchfactor}{4}
\providecommand{\BIBentryALTinterwordspacing}{\spaceskip=\fontdimen2\font plus
\BIBentryALTinterwordstretchfactor\fontdimen3\font minus
  \fontdimen4\font\relax}
\providecommand{\BIBforeignlanguage}[2]{{%
\expandafter\ifx\csname l@#1\endcsname\relax
\typeout{** WARNING: IEEEtran.bst: No hyphenation pattern has been}%
\typeout{** loaded for the language `#1'. Using the pattern for}%
\typeout{** the default language instead.}%
\else
\language=\csname l@#1\endcsname
\fi
#2}}
\providecommand{\BIBdecl}{\relax}
\BIBdecl

\bibitem{Koyama:1995ys}
K.~{Koyama}, R.~{Petre}, E.~V. {Gotthelf}, U.~{Hwang}, M.~{Matsuura},
  M.~{Ozaki}, and S.~S. {Holt}, ``{Evidence for shock acceleration of
  high-energy electrons in the supernova remnant SN1006},'' \emph{\nat}, vol.
  378, pp. 255--258, Nov. 1995.

\bibitem{Kulkarni:1998lh}
S.~R. {Kulkarni}, D.~A. {Frail}, M.~H. {Wieringa}, R.~D. {Ekers}, E.~M.
  {Sadler}, R.~M. {Wark}, J.~L. {Higdon}, E.~S. {Phinney}, and J.~S. {Bloom},
  ``{Radio emission from the unusual supernova 1998bw and its association with
  the {$\gamma$}-ray burst of 25 April 1998},'' \emph{\nat}, vol. 395, pp.
  663--669, Oct. 1998.

\bibitem{Enomoto:2002rt}
R.~{Enomoto}, T.~{Tanimori}, T.~{Naito}, T.~{Yoshida}, S.~{Yanagita},
  M.~{Mori}, P.~G. {Edwards}, A.~{Asahara}, G.~V. {Bicknell}, S.~{Gunji},
  S.~{Hara}, T.~{Hara}, S.~{Hayashi}, C.~{Itoh}, S.~{Kabuki}, F.~{Kajino},
  H.~{Katagiri}, J.~{Kataoka}, A.~{Kawachi}, T.~{Kifune}, H.~{Kubo},
  J.~{Kushida}, S.~{Maeda}, A.~{Maeshiro}, Y.~{Matsubara}, Y.~{Mizumoto},
  M.~{Moriya}, H.~{Muraishi}, Y.~{Muraki}, T.~{Nakase}, K.~{Nishijima},
  M.~{Ohishi}, K.~{Okumura}, J.~R. {Patterson}, K.~{Sakurazawa}, R.~{Suzuki},
  D.~L. {Swaby}, K.~{Takano}, T.~{Takano}, F.~{Tokanai}, K.~{Tsuchiya},
  H.~{Tsunoo}, K.~{Uruma}, A.~{Watanabe}, and T.~{Yoshikoshi}, ``{The
  acceleration of cosmic-ray protons in the supernova remnant RX
  J1713.7-3946},'' \emph{\nat}, vol. 416, pp. 823--826, Apr. 2002.

\bibitem{Bamba:2003vn}
A.~{Bamba}, R.~{Yamazaki}, M.~{Ueno}, and K.~{Koyama}, ``{Small-Scale Structure
  of the SN 1006 Shock with Chandra Observations},'' \emph{\apj}, vol. 589, pp.
  827--837, Jun. 2003.

\bibitem{Piran:1999jt}
T.~{Piran}, ``{Gamma-ray bursts and the fireball model},'' \emph{\physrep},
  vol. 314, pp. 575--667, Jun. 1999.

\bibitem{Medvedev:1999sf}
M.~V. {Medvedev} and A.~{Loeb}, ``{Generation of Magnetic Fields in the
  Relativistic Shock of Gamma-Ray Burst Sources},'' \emph{\apj}, vol. 526, pp.
  697--706, Dec. 1999.

\bibitem{Lyutikov:2003fk}
M.~{Lyutikov} and R.~{Blandford}, ``{Gamma Ray Bursts as Electromagnetic
  Outflows},'' \emph{astro-ph/0312347}, Dec. 2003.

\bibitem{Fox:2006vs}
D.~B. {Fox} and P.~{M{\'e}sz{\'a}ros}, ``{GRB fireball physics: prompt and
  early emission},'' \emph{New Journal of Physics}, vol.~8, pp. 199--+, Sep.
  2006.

\bibitem{Fender:1999fk}
R.~P. {Fender}, S.~T. {Garrington}, D.~J. {McKay}, T.~W.~B. {Muxlow}, G.~G.
  {Pooley}, R.~E. {Spencer}, A.~M. {Stirling}, and E.~B. {Waltman}, ``{MERLIN
  observations of relativistic ejections from GRS 1915+105},'' \emph{\mnras},
  vol. 304, pp. 865--876, Apr. 1999.

\bibitem{Kaiser:2000fj}
C.~R. {Kaiser}, R.~{Sunyaev}, and H.~C. {Spruit}, ``{Internal shock model for
  microquasars},'' \emph{\aap}, vol. 356, pp. 975--988, Apr. 2000.

\bibitem{Jamil:2010lr}
O.~{Jamil}, R.~P. {Fender}, and C.~R. {Kaiser}, ``{iShocks: X-ray binary jets
  with an internal shocks model},'' \emph{\mnras}, vol. 401, pp. 394--404, Jan.
  2010.

\bibitem{Rees:1978zr}
M.~J. {Rees}, ``{The M87 jet - Internal shocks in a plasma beam},''
  \emph{\mnras}, vol. 184, pp. 61P--65P, Sep. 1978.

\bibitem{Spada:2001mz}
M.~{Spada}, G.~{Ghisellini}, D.~{Lazzati}, and A.~{Celotti}, ``{Internal shocks
  in the jets of radio-loud quasars},'' \emph{\mnras}, vol. 325, pp.
  1559--1570, Aug. 2001.

\bibitem{Sahayanathan:2005uq}
S.~{Sahayanathan} and R.~{Misra}, ``{Interpretation of the Radio/X-Ray Knots of
  AGN Jets within the Internal Shock Model Framework},'' \emph{\apj}, vol. 628,
  pp. 611--616, Aug. 2005.

\bibitem{Drury:1983sf}
L.~O. {Drury}, ``{An introduction to the theory of diffusive shock acceleration
  of energetic particles in tenuous plasmas},'' \emph{Reports on Progress in
  Physics}, vol.~46, pp. 973--1027, Aug. 1983.

\bibitem{Bret:2009ec}
A.~{Bret}, ``{Weibel, Two-Stream, Filamentation, Oblique, Bell, Buneman...Which
  One Grows Faster?}'' \emph{\apj}, vol. 699, pp. 990--1003, Jul. 2009.

\bibitem{Lemoine:2009kg}
M.~{Lemoine} and G.~{Pelletier}, ``{On electromagnetic instabilities at
  ultra-relativistic shock waves},'' \emph{\mnras}, pp. 1732--+, Nov. 2009.

\bibitem{Weibel:1959it}
E.~S. {Weibel}, ``{Spontaneously Growing Transverse Waves in a Plasma Due to an
  Anisotropic Velocity Distribution},'' \emph{Physical Review Letters}, vol.~2,
  pp. 83--84, Feb. 1959.

\bibitem{Morse:1971}
R.~L. {Morse} and C.~W. {Nielson}, ``{Numerical Simulation of the Weibel
  Instability in One and Two Dimensions},'' \emph{Physics of Fluids}, vol.~14,
  pp. 830--840, 1971.

\bibitem{Davidson:1972}
R.~C. {Davidson}, D.~A. {Hammer}, I.~{Haber}, and C.~E. {Wagner}, ``{Nonlinear
  Development of Electromagnetic Instabilities in Anisotropic Plasmas},''
  \emph{Physics of Fluids}, vol.~15, pp. 317--333, 1972.

\bibitem{Lee:1973}
R.~{Lee} and M.~{Lampe}, ``{Electromagnetic Instabilities, Filamentation, and
  Focusing of Relativistic Electron Beams},'' \emph{Physical Review Letters},
  vol.~31, pp. 1390--1393, 1973.

\bibitem{Molvig:1975}
K.~{Molvig}, ``{Filamentary Instability of a Relativistic Electron Beam},''
  \emph{Physical Review Letters}, vol.~35, pp. 1504--1507, 1975.

\bibitem{Sagdeev:1979ph}
R.~Z. {Sagdeev}, ``{The 1976 Oppenheimer lectures: Critical problems in plasma
  astrophysics. II. Singular layers and reconnection},'' \emph{Reviews of
  Modern Physics}, vol.~51, pp. 11--20, Jan. 1979.

\bibitem{Bret:2004dn}
A.~{Bret}, M.-C. {Firpo}, and C.~{Deutsch}, ``{Collective electromagnetic modes
  for beam-plasma interaction in the whole k space},'' \emph{\pre}, vol.~70,
  no.~4, pp. 046\,401--+, Oct. 2004.

\bibitem{Bret:2005hl}
------, ``{Characterization of the Initial Filamentation of a Relativistic
  Electron Beam Passing through a Plasma},'' \emph{Physical Review Letters},
  vol.~94, no.~11, pp. 115\,002--+, Mar. 2005.

\bibitem{Lazar:2009qq}
M.~{Lazar}, A.~{Smolyakov}, R.~{Schlickeiser}, and P.~K. {Shukla}, ``{A
  comparative study of the filamentation and Weibel instabilities and their
  cumulative effect. I. Non-relativistic theory},'' \emph{Journal of Plasma
  Physics}, vol.~75, pp. 19--+, 2009.

\bibitem{Honda:2000}
M.~{Honda}, J.~{Meyer-ter-Vehn}, and A.~{Pukhov}, ``{Two-dimensional
  particle-in-cell simulation for magnetized transport of ultra-high
  relativistic currents in plasma},'' \emph{Physics of Plasmas}, vol.~7, pp.
  1302--1308, 2000.

\bibitem{Silva:2002qc}
L.~O. {Silva}, R.~A. {Fonseca}, J.~W. {Tonge}, W.~B. {Mori}, and J.~M.
  {Dawson}, ``{On the role of the purely transverse Weibel instability in fast
  ignitor scenarios},'' \emph{Physics of Plasmas}, vol.~9, pp. 2458--2461, Jun.
  2002.

\bibitem{Stockem:2008hl}
A.~{Stockem}, M.~E. {Dieckmann}, and R.~{Schlickeiser}, ``{Suppression of the
  filamentation instability by a flow-aligned magnetic field: testing the
  analytic threshold with PIC simulations},'' \emph{Plasma Physics and
  Controlled Fusion}, vol.~50, no.~2, pp. 025\,002--+, Feb. 2008.

\bibitem{Califano:2002ci}
F.~{Califano}, T.~{Cecchi}, and C.~{Chiuderi}, ``{Nonlinear kinetic regime of
  the Weibel instability in an electron-ion plasma},'' \emph{Physics of
  Plasmas}, vol.~9, pp. 451--457, Feb. 2002.

\bibitem{Zenitani:2008qy}
S.~{Zenitani} and M.~{Hesse}, ``{The role of the Weibel instability at the
  reconnection jet front in relativistic pair plasma reconnection},''
  \emph{Physics of Plasmas}, vol.~15, no.~2, pp. 022\,101--+, Feb. 2008.

\bibitem{Swisdak:2008fj}
M.~{Swisdak}, Y.~{Liu}, and J.~F. {Drake}, ``{Development of a Turbulent
  Outflow During Electron-Positron Magnetic Reconnection},'' \emph{\apj}, vol.
  680, pp. 999--1008, Jun. 2008.

\bibitem{Silva:2003rw}
L.~O. {Silva}, R.~A. {Fonseca}, J.~W. {Tonge}, J.~M. {Dawson}, W.~B. {Mori},
  and M.~V. {Medvedev}, ``{Interpenetrating Plasma Shells: Near-equipartition
  Magnetic Field Generation and Nonthermal Particle Acceleration},''
  \emph{\apjl}, vol. 596, pp. L121--L124, Oct. 2003.

\bibitem{Kazimura:1998}
Y.~{Kazimura}, J.~I. {Sakai}, T.~{Neubert}, and S.~V. Bulanov, ``{Generation of
  a small-scale quasi-static magnetic field and fast particles during the
  collision of electron-positron plasma clouds},'' \emph{Astrophysical
  Journal}, vol. 498, pp. L183--L186, 1998.

\bibitem{Spitkovsky:2005hi}
A.~{Spitkovsky}, ``{Simulations of relativistic collisionless shocks: shock
  structure and particle acceleration},'' in \emph{Astrophysical Sources of
  High Energy Particles and Radiation}, ser. American Institute of Physics
  Conference Series, T.~{Bulik}, B.~{Rudak}, and G.~{Madejski}, Eds., vol. 801,
  Nov. 2005, pp. 345--350.

\bibitem{Sironi:2009kt}
L.~{Sironi} and A.~{Spitkovsky}, ``{Particle Acceleration in Relativistic
  Magnetized Collisionless Pair Shocks: Dependence of Shock Acceleration on
  Magnetic Obliquity},'' \emph{\apj}, vol. 698, pp. 1523--1549, Jun. 2009.

\bibitem{Nishikawa:2009}
K.~I. {Nishikawa}, J.~{Niemiec}, P.~E. {Hardee}, M.~{Medvedev}, H.~{Sol},
  Y.~{Mizuno}, B.~{Zhang}, M.~{Pohl}, M.~{Oka}, and D.~H. {Hartmann}, ``{Weibel
  instability and associated strong fields in a fully three-dimensional
  simulation of a relativistic shock},'' \emph{Astrophysical Journal Letters},
  vol. 698, pp. L10--L13, 2009.

\bibitem{Frederiksen:2004rm}
J.~T. {Frederiksen}, C.~B. {Hededal}, T.~{Haugb{\o}lle}, and {\AA}.~{Nordlund},
  ``{Magnetic Field Generation in Collisionless Shocks: Pattern Growth and
  Transport},'' \emph{\apjl}, vol. 608, pp. L13--L16, Jun. 2004.

\bibitem{Spitkovsky:2008Ion}
A.~{Spitkovsky}, ``{On the structure of relativistic collisionless shocks in
  electron-ion plasmas },'' \emph{Astrophysical Journal Letters}, vol. 673, pp.
  L39--L42, 2008.

\bibitem{Hededal:2005zl}
C.~B. {Hededal} and K.-I. {Nishikawa}, ``{The Influence of an Ambient Magnetic
  Field on Relativistic collisionless Plasma Shocks},'' \emph{\apjl}, vol. 623,
  pp. L89--L92, Apr. 2005.

\bibitem{BreGrem:2007}
A.~{Bret}, L.~{Gremillet}, and J.~{Bellido}, ``{How really transverse is the
  filamentation instability?}'' \emph{Physics of Plasmas}, vol.~14, p. 032103,
  2007.

\bibitem{Bessho:1999sf}
N.~{Bessho} and Y.~{Ohsawa}, ``{Electron acceleration to ultrarelativistic
  energies in a collisionless oblique shock wave},'' \emph{Physics of Plasmas},
  vol.~6, pp. 3076--3085, Aug. 1999.

\bibitem{Dieckmann:2008dp}
M.~E. {Dieckmann}, P.~K. {Shukla}, and L.~O.~C. {Drury}, ``{The Formation of a
  Relativistic Partially Electromagnetic Planar Plasma Shock},'' \emph{\apj},
  vol. 675, pp. 586--595, Mar. 2008.

\bibitem{Dawson:1983dz}
J.~M. {Dawson}, ``{Particle simulation of plasmas},'' \emph{Reviews of Modern
  Physics}, vol.~55, pp. 403--447, Apr. 1983.

\bibitem{Buneman:1958}
O.~{Buneman}, ``{Instability, Turbulence, and Conductivity in Current-Carrying
  Plasma},'' \emph{Physical Review Letters}, vol.~1, pp. 8--9, 1958.

\bibitem{Bret:2006rc}
A.~{Bret}, M.~E. {Dieckmann}, and C.~{Deutsch}, ``{Oblique electromagnetic
  instabilities for a hot relativistic beam interacting with a hot and
  magnetized plasma},'' \emph{Physics of Plasmas}, vol.~13, no.~8, pp.
  082\,109--+, Aug. 2006.

\bibitem{Bogdan:1985}
T.~J. {Bogdan} and I.~{Lerche}, ``{Dynamical evolution of large-scale
  two-dimensional, fibril magnetic fields},'' \emph{Astrophysical Journal},
  vol. 296, pp. 719--738, 1985.

\bibitem{Hammer:1970}
D.~A. {Hammer} and N.~{Rostocker}, ``Propagation of high current relativistic
  electron beams,'' \emph{Physics of Fluids}, vol.~13, pp. 1831--1850, 1970.

\bibitem{Mann:1994}
G.~{Mann}, H.~{Luhr}, and W.~{Baumjohann}, ``Statistical analysis of short
  large-amplitude magnetic field structures in the vicinity of the
  quasi-parallel bow shock,'' \emph{Journal of Geophysical Research}, vol.~99,
  pp. 13\,315--13\,323, 1994.

\bibitem{Behlke:2003}
R.~{Behlke}, M.~{Andre}, S.~C. {Buchert}, A.~{Vaivads}, A.~I. {Eriksson}, E.~A.
  {Lucek}, and A.~{Balogh}, ``Multi-point electric field measurements of short
  large-amplitude magnetic structures (slams) at the earth's quasi-parallel bow
  shock,'' \emph{Geophysical Research Letters}, vol.~30, p. 1177, 2003.

\bibitem{Lembege:2004}
K.~{Tsubouchi} and B.~{Lembege}, ``Full particle simulations of short
  large-amplitude magnetic structures (slams) in quasi-parallel shocks,''
  \emph{Journal of Geophysical Research}, vol. 109, p. A02114, 2004.

\bibitem{Dieckmann:2010qf}
M.~E. {Dieckmann}, G.~C. {Murphy}, A.~{Meli}, and L.~O.~C. {Drury},
  ``{Particle-in-cell simulation of a mildly relativistic collision of an
  electron-ion plasma carrying a quasi-parallel magnetic field . Electron
  acceleration and magnetic field amplification at supernova shocks},''
  \emph{\aap}, vol. 509, pp. A89+, Jan. 2010.

\bibitem{Dieckmann:2004}
M.~E. {Dieckmann}, A.~{Ynnerman}, S.~C. {Chapman}, G.~{Rowlands}, and
  N.~{Andersson}, ``Simulating thermal noise,'' \emph{Physica Scripta},
  vol.~69, pp. 456--460, 2004.

\bibitem{Bell:2004eu}
A.~R. {Bell}, ``{Turbulent amplification of magnetic field and diffusive shock
  acceleration of cosmic rays},'' \emph{\mnras}, vol. 353, pp. 550--558, Sep.
  2004.

\end{thebibliography}
\end{document}